# Sample to sample fluctuations in fragmentation and agglomeration processes

Piero Olla

ISIAtA, Area Ricerca CNR
Università degli Studi Lecce
Lecce, 73100 Italy

**Abstract**

The fluctuation dynamics in steady state fragmentation and aggregation processes is analyzed from a kinetic point of view. The fluctuations $\Delta n(x,t)$ in the number of particles of size between $x$ and $x+\Delta x$, in a sample with average occupation number $\Delta \bar{N}(x)$, appear to behave like sums of uncorrelated events: $<\Delta n(x,t)^2> \propto \Delta \bar{N}(x)$. The implications of this fact for the applicability of a mean field description, and possible relation with intermittency phenomena in three dimensional turbulence are discussed.

The analysis of fragmentation and agglomeration processes focuses usually on the determination of the distribution $\bar N(x,t)$ for the particle size $x$, which gives the mean field approximation for the problem. Once an explicit expression for the transition probability $\alpha$ for the process is available, the distribution $\bar N$ can be obtained from a Boltzmann-like equation, which is linear for binary fragmentation (each break-up event results in only two fragments) [1-2]:

$$\partial_t \bar N(x,t) = 2\int_x^\infty dy\,\alpha(y\to x, y-x)\bar N(y,t) - \bar N(x,t)\int_0^x dy\,\alpha(x\to y, x-y), \qquad (1)$$

while, for agglomeration from binary collisions, one has the Smoluchowski equation [3]:

$$\partial_t \bar N(x,t) = \frac{1}{2}\int_0^x dy\,\alpha(y, x-y \to x)\bar N(y,t)\bar N(x-y,t)$$
$$-\bar N(x,t)\int_0^\infty dy\,\alpha(x, y \to x+y)\bar N(y,t). \qquad (2)$$

Clearly, while Eqn. (1) is exact, Eqn. (2) is an approximation, which neglects correlations and is based usually on assumptions of diluteness [4].

The field of application of Eqns. (1-2), is clearly very wide (see for instance [5] and [6] and references therein). The possibility of using a mean field description like the one provided by Eqns. (1) and (2) depends on whether fluctuations can be considered small. In the case of agglomeration, this affects the validity of the approximations leading to Eqn. (2), since large fluctuations mean that correlations in the form: $<N(x,t)N(y,t)>$, with $<>$ indicating ensemble average, can dominate over the factorized terms $\bar N(x,t)\bar N(y,t) \equiv <N(x,t)><N(y,t)>$.

The role of fluctuations in fragmentation was examined recently By Esipov et Al., in the case of break-up probabilities in the form: $\alpha(x \to y, x-y) = x^\beta$ [7]. Their analysis showed a situation of dominance of the fluctuations down the cascade in the shattering range [6], corresponding to fragmentation accelerating down the cascade in such a way that an initial object disintegrates into zero size fragments in a finite time. The relevance to intermittency issues in turbulence modeling (see e.g. [8]) is obvious. However, the reason why a transition to dominant fluctuations occurs in this way is not yet clear.

The purpose of this letter is to clarify this issue, at least in situations of constant mass current, and to extend the analysis of [7] to agglomeration processes. Despite the differences between fragmentation and agglomeration processes, it will appear that the mechanism for the transition to a situation of fluctuation dominance, is the same in both. While in [7], the approach was essentially a statistical mechanical one, based on an analysis in "$\Gamma$-space" of the probability functional $\mathcal{P}[N]$, the approach here will be to study directly the "$\mu$-space" kinetic equations for the $N(x,t)$, whose mean field approximation is given by Eqns. (1) and (2).

Fluctuations are analyzed identifying a set of mother particles, and defining a sample, as the subset of fragments (or agglomerates), which originate from them. The mother particles can be characterized for instance as the largest (in the case of fragmentation) or smallest (in the case of agglomeration) particles, which are present at any time, in a prescribed volume element $V$ of the system. The volume $V$ is chosen large enough to guarantee a sufficient sample, which, in the case of coalescence, is necessary also to minimize interactions with particles coming from the outside of $V$. In this way it will be possible also to consider $n(x,t) = N(x,t) - \bar N(x)$ as a spatial fluctuation.

*Binary fragmentation*

The kinetic equation for a fragmentation process can be written in the form, after discretizing in time and particle size:

$$N(x,t+1) = N(x,t) + 2\sum_{y=x+1}^\infty W(y,x;t) - 2\sum_{y=1}^{x-1} W(x,y;t), \qquad (3)$$

where $W(x,y;t) = \sum_{j=1}^{N(x,t)} w(x \to y, x - y, t; j)$, with $w(x \to y, x - y, t; j)$ a stochastic variable that is 1 or 0, depending on whether the $j$-th particle of size $x$ undergoes or not the breaking process: $x \to y, x - y$. If the break-up events occur independently, the probability that the sums $W(x,y;t)$ take certain values, will be that of distributing $W(x,y;t)$ particles, out of $N(x,t)$, in slots $y = 1, ...x - 1$, with: $W(x,y;t) = W(x, x - y; t)$. This is a multinomial distribution that, for $\bar{N}(x) \gg 1$, is approximated by:

$$\mathcal{P}[W] \simeq \prod_x \prod_{y=1}^{x'} \exp\left( - \frac{\Delta W(x, y; t)^2}{2\alpha(x, y)\bar{N}(x)} \right) \delta_{W(x,y),W(y,x)} \qquad (4)$$

where $\Delta W(x, y; t) = W(x, y; t) - \alpha(x, y)\bar{N}(x)$, $x' = \frac{x}{2}$ for $x$ even and $x' = \frac{x+1}{2}$ for odd $x$, while $\alpha(x, y) \equiv \alpha(x \to y, x - y)$ is the probability of an individual breaking event. Averaging Eqn. (3) over the distribution given by Eqn. (4) and taking the continuum limit, gives immediately Eqn. (1), while the fluctuations obey the stochastic (linear) integral equation:

$$\partial_t n(x,t) - 2 \int_x^\infty dy \, \alpha(y,x) n(y,t) + n(x,t) \int_0^x dy \, \alpha(x,y) = \xi(x,t) \qquad (5)$$

where:

$$\xi(x,t) = 2 \int_x^\infty dy \, \Delta W(y,x;t) - \int_0^x dy \, \Delta W(x,y;t), \qquad (6)$$

plays the role of external noise term. Focusing on the case of homogeneous breaking probabilities: $\alpha(x,y) = x^\beta f(y/x)$ with $\lim_{z \to 0} zf(z) = 0$, one obtains a stationary solution of Eqn. (1), requiring the mass current $J(x) = 2 \int_x^\infty dy \int_0^x dz \, \alpha(y,z) z \bar{N}(y)$ to be $x$-independent. From power counting: $\bar{N}(x) \simeq x^{-\gamma}$, with $\gamma = 3 + \beta$.

Given the statistics for $\Delta W$, the correlation for the external noise $\xi$ can be calculated and one finds that it contains a singular piece: $< \xi(x,t)\xi(x',t') >_s \simeq \delta(t - t')\delta(x - x') x^{-2}$. The correlation for $n$, therefore, receives itself a singular contribution: $< n(x,t)n(x',t') >_s$, whose dynamics is dominated by the balance between $< \xi(x,t)\xi(x',t') >_s$, and the loss term $n(x,t) \int_0^x dy \, \alpha(x,y)$ in eqn. (5). One obtains therefore:

$$< n(x,t)n(x',t') >_s \propto \delta(x - x')\bar{N}(x) \exp(-\Gamma(x)|t - t'|), \qquad (7)$$

with $\Gamma(x) = x^{\beta+1} \int_0^1 dz \, f(z)$. This result implies that the fluctuations $\Delta n(x,t)$ in the number of particles, in a small interval of size $\Delta x$, behave like the fluctuations in a sample of statistically independent events: $< \Delta n(x)^2 > \simeq \Delta \bar{N}(x)$. From Eqn. (7), fluctuations become dominant down the cascade: $\frac{<\Delta n(x)^2>}{\Delta \bar{N}(x)^2} \to \infty$ as $x \to 0$, when $\Delta \bar{N}(x \to 0) = 0$, i.e. for $\beta < -2$; this coincides with the result obtained by Esipov et Al. [7] in the case of a Yule-Ferrin cascade [9].

These results are contingent upon neglecting any correlations among the break-up events. In particular, if one admits a finite correlation time in the fragmentation rate fluctuation $\xi(x,t)$, (which is especially important for turbulence modeling) different mechanisms for fluctuations generation become possible. The issue is already apparent, in as simple a turbulence model as a Yule-Ferrin cascade, in which each eddy splits into two equal fragments, conserving eddy energy and volume. From Kolmogorov scaling [10], one has for the energy of the eddies, at the $m$-th step in the cascade: $E_m = N_m x_m \sim l_m^{\frac{2}{3}} = 2^{-\frac{2}{3}m}$, where $N_m \equiv \Delta N(x_m)$, and $x_m = 2^{-m}$ is the energy of the individual eddy in the $m$-th generation, with linear size $l_m = 2^{-\frac{m}{3}}$. This fixes also the break-up probability: $\alpha_m \equiv \alpha(x_m, x_{m+1}) = E_m^{-1}$. If, following [11], the finite correlation time of the break-up process is kept into account, the amplitude of the external noise in Eqn. (5) is corrected by a factor $\alpha_m^{-1}$, the result: $<\Delta n(x)^2>\sim \bar N(x)$ ceases to be valid, and one obtains the equation for $\phi_m = \bar N_m^{-1} n_m$:

$$\partial_t \phi_m + \alpha_m \partial_m \phi_m = \alpha_m \partial_m (\alpha_m^{-\frac{1}{2}} \xi_m), \tag{8}$$

where $\partial_m$ is a finite difference and $<\xi_m(t)\xi_{m'}(t')> = \delta_{mm'}\delta(t-t')$. Solution of this equation gives then the result: $<\phi_m^2>\sim 1 + m(\partial_m \log \alpha_m)^2$, in which the first term comes from power counting, that is the fluctuation level that would be expected from Kolmogorov scaling, while the second, associated with production of intermittency, comes explicitly from the factor $\alpha_m^{-\frac{1}{2}}$ in front of $\xi$, i.e. from the finiteness of the correlation time [12].

*Coalescence processes*

In this case, the starting kinetic equation, can be written in the form, after discretization:

$$N(x, t+1) = N(x,t) + \frac{1}{2}\sum_{y=1}^{x-1} W(y, x-y; t) - \sum_{y=1}^{\infty} W(x,y;t), \tag{9}$$

where $W(x,y;t) = \sum_{j_x j_y} w(x,y \to x+y, t; j_x, j_y)$, with $w$ equal to one, if particles $j_x$ and $j_y$ coalesce into an object of mass $x+y$, and to zero if they do not. The particles are supposed distributed at random over scales below that of the mean free path. The reason for this assumption is to focus on production of fluctuations at macroscopic scales, rather than from problems in the process of coarse graining.

Dividing the volume $V$, into $K$ slots of size much larger than that of the particles, the probability of having $W_2(x,y)$ slots with two particles, will be a multinomial distribution, that in the limit: $K \gg \bar N(x), \bar N(y) \gg W_2(x,y) \gg 1$, will be approximated by a multivariate gaussian. Notice that the inequality $K \gg \bar N(x)$ is nothing else than the diluteness condition, while $W_2(x,y) \gg 1$ is the large sample condition necessary for coarse graining. The probability of having $W(x,y)$ coalescence events will be a gaussian as well in the hypotheses: $1 \ll W(x,y) \ll W_2(x,y)$, so that, analogously to Eqn. (4), it is possible to write:

$$\mathcal{P}[W] \simeq \prod_{y \geq x} \exp\Big(-\frac{\Delta W(x,y;t)^2}{2\alpha(x,y)\bar N(x)\bar N(y)}\Big) \delta_{W(x,y),W(y,x)} \tag{10}$$

where: $\alpha(x,y) \equiv \alpha(x,y \to x+y) = V^{-1}\sigma(x,y)v(x,y)$, with $\sigma$ the coalescence cross section, and $v$ an effective velocity; $[K\alpha(x,y)$ is the coalescence probability for two particles in the same slot]. In the continuum limit, the average of Eqn. (9) gives immediately Eqn. (2), while the fluctuations are governed by the stochastic equation:

$$\partial_t n(x,t) + n(x,t)\int_0^\infty dy\, \alpha(x,y)\bar N(y) - \xi(x,t)$$

$$= \int_0^x dy\, \alpha(x-y,y)\bar N(x-y)n(y,t) - \bar N(x)\int_0^\infty dy\, \alpha(x,y)n(y,t), \tag{11}$$

where:

$$\xi(x,t) = \frac{1}{2}\int_0^x dy\, \Delta W(x-y,y;t) - \int_0^\infty dy\, \Delta W(x,y;t). \tag{12}$$

Focusing again on power law kernels: $\alpha(ax,ay) = a^\beta \alpha(x,y)$, with: $\alpha(x,y)|_{x \gg y} \simeq x^\delta y^{\beta-\delta}$, there will be a constant current solution: $\bar{N}(x) = x^{-\gamma}$ with $\gamma = \frac{3+\beta}{2}$, which will be valid if the infinities in Eqn. (2) cancel one another, i.e. when: $\delta < \frac{1+\beta}{2}$.

The noise correlation $< \xi(x,t)\xi(x',t') >$ is obtained from Eqns. (10) and (12) and, as in the case of fragmentation, contains a singular piece proportional to $\delta(x-x')$:

$$< \xi(x,t)\xi(x',t') > = \delta(t-t')[\Xi_S(x)\delta(x-x') + \Xi(x,x')], \qquad (13)$$

where: $\Xi_S(x) = \frac{1}{2}\int_0^x dy \alpha(x-y,y)\bar{N}(x-y)\bar{N}(y) + \int_0^\infty dy \alpha(x,y)\bar{N}(x)\bar{N}(y)$ and $\Xi(x,x') = \alpha(x',x-x')\bar{N}(x')\bar{N}(x-x')$ for $x > x'$. This leads to the expression:

$$< n(x,t)n(x',t') >_S = \frac{3}{4}\delta(x-x')\bar{N}(x)\exp(-\Gamma(x)|t-t'|), \qquad (14)$$

where: $\Gamma(x) = \int_0^\infty dy \alpha(x,y)\bar{N}(y)$. There are now two possibilities:

A: if $\delta < \frac{\beta-1}{2}$, the correlation time $\Gamma(x)$ is finite, and the same situation of fragmentation, with: $< \Delta n(x)^2 > \sim \Delta \bar{N}(x)$, arises. Systems in which particle interactions are dominated by fluid mechanics are likely to behave like this [13].

B: if $\frac{1+\beta}{2} > \delta > \frac{\beta-1}{2}$, the correlation time decreases with $x$, like a power of $x_0/x$, with $x_0$ the smallest particles in the systems. The final outcome, at large $x/x_0$, is one of singular fluctuations behaving like a white noise in time, with vanishing amplitude $\sim \Gamma(x)^{-1}$. The kernel for Brownian coagulation: $\alpha(x,y) = \frac{x^{-1/3}+y^{-1/3}}{x^{1/3}+y^{1/3}}$ [3,14], and the product kernel: $\alpha(x,y) = (xy)^\delta$, both lead to this behavior.

Physically, the situation in case (B) is a consequence of the process being dominated by deposition of large numbers of the smallest particles. This leads to a steady growth in the mass of each agglomerate, with fluctuations vanishing, as the size of the deposited particles and their number, go respectively to zero and to infinity.

The non-singular part of the fluctuations becomes important if one wants to verify the validity of Eqn. (2). In case (B), it can be shown that the non-singular correlations: $< n(x,t)n(x',t') >_{NS}$ are continuous in $x$ and $x'$, so that, if $\delta < \frac{1+\beta}{2}$, the divergences at $y = 0$ in the integrals in Eqn. (11) cancel one another. In case (A), problems with divergences are absent to start with. Hence, power counting can be used in both cases without any problem, and: $< n(x,t)n(x',t) >_{NS} \sim x^{-1}\bar{N}(x)g(x/x')$. The condition for correlations being negligible: $\frac{<n(x)n(y)>}{\bar{N}(x)\bar{N}(y)} \to 0$ as $x \to \infty$, becomes therefore: $\bar{N}(x) \sim x^{-\gamma}$, with $\gamma < 1$, i.e.: $\Delta \bar{N}(x \to \infty) = \infty$ [15]. This is nothing else than the requirement of preservation of diluteness and large sample size as $x \to \infty$.

To conclude: fluctuations in both fragmentation and coalescence processes satisfy the rule $< \Delta n(x)^2 > \sim \Delta \bar{N}(x)$ characteristic of uncorrelated processes. The origin of the non self-averaging properties of shattering systems, already observed in [7], and of agglomerating systems, for $\beta > -1$ [see definition after Eqn. (12)], appears to reside only in the decrease in the direction of the processes, of $\Delta \bar{N}(x) \sim x\bar{N}(x)$. Also the vanishing of $\frac{<\Delta n^2(x)>}{\Delta \bar{N}(x)}$, for fixed $\Delta x$ and $x \to \infty$, in the case of deposition dominated coalescence, is a manifestation of the lack of particle correlations, and fluctuations are killed in this case simply by the large number of deposited particles.

The first reason behind these results, is the independence of the individual breaking or coalescence events, and the absence of memory in the solutions of the kinetic equations. These are very common properties in the case of fragmentation. In the case of agglomeration, however, assumptions on the absence of ordered motions at small scales become necessary, which are natural in cases like e.g. that of Brownian coagulation, but are more difficult to justify in general.

The second reason for lack of particle correlation, is the Markovian nature of the processes considered. In the case of fragmentation, in particular, violations of the result: $<\Delta n^2> \sim \Delta \bar{N}$ are obtained, as soon as the correlation time for the breaking events becomes non-zero. This effect is what lies behind the generation of intermittency in turbulence models of the type recently studied in [11-12].

**Aknowldedgments**: I would like to thank Gianluigi Zanetti of CRS4, where part of this research was carried on, for hospitality.